\documentclass[prd,onecolumn,notitlepage,nofootinbib,superscriptaddress,showpacs,showkeys]{revtex4-2}

\usepackage{amsfonts}
\usepackage{graphicx}
\usepackage{dcolumn}
\usepackage{amsmath}
\usepackage{amssymb}
\usepackage[T1]{fontenc}
\usepackage[utf8]{inputenc}
\usepackage{esint}
\usepackage[brazil,english]{babel}
\usepackage[usenames,dvipsnames]{color}
\usepackage{longtable}
\usepackage{ulem}
\usepackage{nicefrac}
\usepackage{bm}
\usepackage{tikz}
\usepackage{braket}
\usepackage{graphicx}

\begin{document}

\title{Dynamical black holes and accretion-induced backreaction}

\author{Thiago de L. Campos}
\email{thiagocampos@usp.br}
\affiliation{Instituto de F\'{\i}sica, Universidade de S\~{a}o Paulo, Caixa Postal 66318,
05315-970, S\~{a}o Paulo-SP, Brazil}

\author{C. Molina}
\email{cmolina@usp.br}
\affiliation{Escola de Artes, Ci\^{e}ncias e Humanidades, Universidade de S\~{a}o Paulo, Avenida Arlindo Bettio 1000, CEP 03828-000, S\~{a}o Paulo-SP, Brazil}

\author{Mario C. Baldiotti}
\email{baldiotti@uel.br}
\affiliation{Departamento de F\'{\i}sica, Universidade Estadual de Londrina, CEP
86051-990, Londrina-PR, Brazil}

\begin{abstract}
We investigate the evolution of future trapping horizons through the dynamics of the Misner-Sharp mass using ingoing Eddington-Finkelstein coordinates. 
Our analysis shows that an integral formulation of Hayward's first law governs much of the evolution of general spherically symmetric spacetimes. 
To account for the accretion backreaction, we consider a near-horizon approximation, yielding first-order corrections of a Vaidya-dark energy form. 
We further propose a systematic perturbative scheme to study these effects for an arbitrary background.
As an application, we analyze an accreting Reissner-Nordström black hole and demonstrate the horizon shifts that are produced. Finally, we compute accretion-induced corrections to an extremal configuration. It is shown that momentum influx and energy density produce distinct effects: the former forces the splitting of the extremal horizon, while the latter induces significant displacements in its position, computed up to first-order perturbative corrections.
These results highlight how different components of the stress-energy tensor significantly affect horizon geometry, with potential implications for broader areas of research, including black-hole thermodynamics.
\end{abstract}

\maketitle

\section{Introduction}
\label{sec:int} 

Black holes are inherently dynamical systems that interact with their environment through accretion or Hawking evaporation.
As horizons evolve and spacetime departs from equilibrium, defining and studying black holes in dynamical regimes requires moving beyond traditional tools like event horizons and Arnowitt-Deser-Misner (ADM)/Komar mass~\cite{krishnan2008fundamental, jaramillo2010mass}.
Quasi-local approaches, such as trapping horizons~\cite{hayward1994general} and the Misner--Sharp mass~\cite{misner1964relativistic}, have emerged as essential counterparts in the descriptions of horizon dynamics and  gravitational energy without relying on particular asymptotic structures~\cite{hernandez1966observer,ashtekar2004isolated,abreu2010kodama,booth2005black,fodor1996surface,nielsen2008dynamical,nielsen2009black,nielsen2009spherically, faraoni2018embedding}, and~avoiding teleological definitions~\cite{visser2014physical}. 
Spherically symmetric spacetimes provide an ideal testing ground for these methods, simplifying the analysis of quasi-local quantities while preserving much of the essential phenomenology of gravitational~dynamics.

While many exact solutions of the Einstein field equations exist, remarkably few describe dynamically interacting black holes with exact treatment for accretion or evaporation.
Notable examples include the McVittie spacetime (and its generalizations) for cosmological black holes~\cite{McVittie:1933zz,carrera,Lake:2011ni,Faraoni:2014nba,daSilva:2015mja,Ruiz:2020yye}, the~Thakurta solution~\cite{Thakurta}, and~the Vaidya metric~\cite{vaidya1951gravitational, vaidya1953newtonian}, which later extended to include the cosmological constant~\cite{mallett1985radiating,campos2024}.
The scarcity of exact solutions for dynamical black holes makes perturbative and approximate approaches indispensable. These methods are crucial for modeling realistic scenarios, such as the backreaction of accreting matter on the metric tensor. For~instance, the~perturbative framework developed by Babichev, Dokuchaev, and~Eroshenko~\cite{babichev2012backreaction} generates first-order metric corrections arising from the energy--momentum tensor in the test-fluid approximation, with~the leading term in a Vaidya-like~form.

As we will show, the~ingoing Eddington--Finkelstein coordinates not only avoid coordinate singularities typical of black holes but also reveal the dynamics of the Misner--Sharp mass in an intuitive way. This is a feature that is largely exploited in our work to analyze general aspects of the evolution of trapping horizons beyond approximate results.
To develop effective descriptions within this framework, with~this chart accommodating future outer (FOTH) and inner (FITH) trapping horizons, we first employ an approximation scheme to simplify the line element near a trapping horizon, yielding particularly tractable results for perfect fluids.
Building on this, we implement a systematic perturbative expansion that goes beyond first-order corrections of matter-energy influx.
A comprehensive analysis of Reissner--Nordstr\"om black holes is carried out, with~special attention given to the extremal~case.

This paper is organized as follows. 
In Section~\ref{sec:dyn} the general spherically symmetric framework is presented. The~dynamics of the Misner--Sharp mass were investigated, with~a connection to Hayward's first law established.
We show in Section~\ref{sec:approx} that the metric admits a suitable approximation in the neighborhood of any 2-sphere (here called ``Vaidya-dark energy''). We implement this approximation scheme specifically for perfect fluids near the horizon. 
In Section~\ref{sec:pert}, we introduce our perturbative approach to accretion, systematically examining its effects on trapping horizon displacements.
The presence of inner horizons is analyzed in Section~\ref{sec:FITH}, with~particular emphasis on how repulsive corrections in the Misner--Sharp mass can eliminate a small FITH of a Reissner--Nordstr\"om black hole.
We conclude with a discussion and outlook in Section~\ref{sec:final}, with Appendix~\ref{app} providing a concise review of future trapping horizon characterization. In~the present work, we use signature {$(- + + +)$}. 


\section{Dynamics of Spherically Symmetric~Spacetimes}
\label{sec:dyn}

\subsection{General~Setup} \label{sec:dyn gen}

This work focuses on a quasi-local approach in spherically symmetric spacetimes. In~this framework, the~Misner--Sharp mass ($M_\text{MS})$, defined as 
\begin{equation}
	M_\text{MS} \equiv \frac{r}{2} ( 1 - g^{\mu\gamma} \partial_\mu r \partial_\gamma r  )  ~,
\end{equation}
can be interpreted as the energy of the system inside a 2-sphere of area $4\pi r^2$ ($r$ denoted as the ``areal radius'').

To study black holes while avoiding coordinate singularities, one can use ingoing Eddington--Finkelstein coordinates $(v,r,\theta,\phi)\,$, where $v$ is the so-called ``advanced time'' (actually a null coordinate). In~this chart, a~general spherically symmetric spacetime is described by the line element~\cite{bardeen1981black}:
\begin{equation}
	ds^2 = - e^{2\lambda(v,r)} \left( 1 - \frac{2M_\text{MS}}{r} \right) dv^2 + 2 e^{\lambda(v,r)} dv dr + r^2 d\Omega^2 ~, ~~ M_\text{MS} \equiv M_\text{MS}(v,r) ~.
	\label{general_metric}
\end{equation}

It is straightforward to calculate the Einstein tensor $G_\mu{}^\gamma$, whose relevant components for this work are
\begin{equation}
	G_v{}^v = - \frac{2 \partial_r M_\text{MS}}{r^2} ~, 
	~~
	G_v{}^r = \frac{2 \partial_v M_\text{MS}}{r^2} ~,
	~~
	G_r{}^v = \frac{2 e^{-\lambda}\partial_r \lambda}{r} ~.
	\label{G}
\end{equation}
The Einstein equation is
\begin{equation}
	G_\mu{}^\gamma = 8\pi \left( T_\mu{}^\gamma - \rho_\Lambda \delta_\mu{}^\gamma \right) ~, 
	~~ G_\mu{}^\gamma \equiv G_\mu{}^\gamma(v,r) ~, 
	~~
	T_\mu{}^\gamma \equiv T_\mu{}^\gamma(v,r) ~, \label{Einst}
\end{equation}
where $\rho_\Lambda \equiv \nicefrac{\Lambda}{8 \pi}$ denotes the energy density of the vacuum.
From the first two components in Equation~\eqref{G}, we obtain
\begin{equation}
	\partial_r M_\text{MS} = - 4 \pi r^2 \left( T_v{}^v - \rho_\Lambda \right) ~, ~~ 
	\partial_v M_\text{MS} = 4 \pi r^2 T_v{}^r~. \label{M'}
\end{equation}
The third component in Equation~\eqref{G} yields
\begin{equation}
	\partial_r e^{-\lambda} = \frac{r}{2} G_r{}^v~,
\end{equation}
which can be integrated to
\begin{equation}
	e^{-\lambda(v,r)} = 4\pi \int^r_{r_0} T_r{}^{v}(v,r')r' \ dr' + e^{-\lambda(v,r_0)}~. \label{e^lambda}
\end{equation}

Under the coordinate transformation $dv \longrightarrow e^{-\lambda(v,r_0)} dv$, the~general line element~\eqref{general_metric} reads 
\begin{equation}
	ds^2 = - e^{2\lambda(v,r) - 2\lambda(v,r_0)} \left( 1 - \frac{2 M_\text{MS}}{r} \right) dv^2 + 2 e^{\lambda(v,r)-\lambda(v,r_0)} dv dr + r^2 d\Omega^2~.
	\label{vr vanish}
\end{equation}
It is apparent that the general metric, written in the form~\eqref{vr vanish}, can be approximated by the Vaidya metric sufficiently close to $r=r_0$. At~this stage, $r_0$ remains an arbitrary~parameter.

\subsection{Dynamics of the Misner--Sharp~Mass} \label{sec:dyn dyn}

To analyze the evolution of the system, we express the Misner--Sharp mass in terms of convenient dynamical variables, using the differential equations in~\eqref{M'}, as
\begin{equation}
	M_\text{MS}(v,r) - M_\text{MS}(0, r_0) = \int^v_0 \mathcal{A}(v',r_0) \ dv' + \int_{r_0}^r \mathcal{B}(v,r') \ dr'~.
	\label{general Misner--Sharp mass}
\end{equation}
The flux and energy functions $\mathcal{A}$ and $\mathcal{B}$ are defined as
\begin{equation}
	\mathcal{A} \equiv 4 \pi r^2 T_v{}^r~, ~~
	\mathcal{B} \equiv - 4 \pi r^2 \left( T_v{}^v - \rho_\Lambda \right) ~.
\end{equation}
These quantities have clear physical interpretations. The~term $\mathcal{A}$ generates the radial energy flux through a sphere of radius $r$ at a fixed $v$, while $\mathcal{B}$ gives the effective energy density within that sphere (including the cosmological constant contribution). In~Equation~\eqref{general Misner--Sharp mass}, $r_{0}$ serves as an arbitrary reference radius that sets the initial condition for the integration.
It follows that the evolution of the Misner--Sharp mass in Equation~\eqref{general Misner--Sharp mass} admits the following physical interpretation. For~any $v>0$ and $r>r_{0}$, the~mass consists of three contributions:
\begin{itemize}
	
	\item the initial mass at the reference radius $r_0$;
	
	\item the accumulated energy flux through the 2-sphere at $r_0$ from $v' = 0$ to $v' = v$;
	
	\item the final energy contained between the 2-spheres at $r_0$ and $r$.
\end{itemize}

We observe that the Misner--Sharp mass at the reference radius $r_{0}$ takes the form
\begin{equation}
	M_\text{MS} (v,r_0) = M_\text{MS}(0, r_0) + \int^v_0 \mathcal{A}(v',r_0) \ dv'~, \label{MS mass (v,r0)}
\end{equation}
provided that $\mathcal{B}$ remains finite. The~boundness of  $\mathcal{B}$ is taken as a physically reasonable assumption for the energy density. Result~\eqref{MS mass (v,r0)} shows that at $r=r_{0}$, the~mass evolves purely as a function of $v$, representing an effective Vaidya solution with initial mass $M_\text{MS}(0,r_0)$.

Let us assume that our system of interest is a black hole with a singularity at $r \rightarrow 0$.
Since $r_0$ is arbitrary in our treatment, we can take the limit $r_0 \rightarrow 0$ in \eqref{MS mass (v,r0)}:
\begin{equation}
	M (v) \equiv  \lim_{r_0 \rightarrow 0} M_\text{MS}(0, r_0) + \int^v_0 \lim_{r_0 \rightarrow 0} \mathcal{A}(v',r_0) \ dv'~.
	\label{mass of the singularity}
\end{equation}
Therefore, we interpret the mass of the singularity itself as $M(v)$ in Equation~\eqref{mass of the singularity}, which adds its initial mass to all the influx of~energy.

Building on Equation~\eqref{general Misner--Sharp mass}, we define the total energy of the black hole at time $v$ as the corresponding Misner--Sharp mass:
\begin{equation}
	M_\text{MS} \left(v, r_H(v) \right) \equiv M(v) + \lim_{r_0 \rightarrow 0} \int_{r_0}^{r_H(v)} \mathcal{B} (v,r') \ dr'~,
	\label{M_BH} 
\end{equation}
where $r_H(v)$ is the position of the black-hole boundary. 
This surface constitutes a future trapping horizon, which is the closure of a hypersurface foliated by 2-dimensional surfaces satisfying
\begin{equation}
\theta_l = 0 ~ , ~~ \theta_n <0 ~ . 
\label{def th}
\end{equation}
In Equation~\eqref{def th}, $\theta_l$ and $\theta_n$ are the expansion scalars associated with the outgoing ($l^\mu$) and ingoing ($n^\nu$) congruences, respectively.
For its full characterization further conditions must be analyzed (see Appendix~\ref{app} for a short exposition on this subject). We denote a future outer trapping horizon (FOTH) as $r=r_+ \,$, a~future inner trapping horizon (FITH) as $r=r_- \,$, and~if it is not specified, we use $r_H$ for the horizon~radius. 

Similarly to Equation~\eqref{M_BH}, we obtain the Misner--Sharp mass decomposition:
\begin{equation}
	M_\text{MS}(v,r) = M(v) + \lim_{r_0 \rightarrow 0} \int_{r_0}^{r} \mathcal{B} (v,r') \ dr'~, \label{MS mass (v,r)}
\end{equation}
where $M(v)$ represents the singularity's time-dependent mass.
Note that the energy--momentum tensor only determines variations in the Misner--Sharp mass since there is a free integration constant [$M_\text{MS}(0,r_0)$], which must be specified in other~ways.  

The presented developments are well contextualized in the so-called Hayward thermodynamics~\cite{hayward1994general, hayward1996gravitational, hayward1998unified, hayward2009local}, which established a consistent thermodynamic description for evolving black holes and wormholes in non-stationary, spherically symmetric spacetimes. Within~this formalism, black holes are defined by their trapping~horizons.

This framework allows deeper insights into the dynamics of the Misner--Sharp mass, especially in non-stationary and strongly curved regimes. For~instance, $M_\text{MS}$ can be interpreted as the conserved quantity associated with the conserved current $J^\mu \equiv G^{\mu\gamma}\text{K}_\gamma$ \cite{faraonicosmological}, where $\text{K}^\mu$ is the Kodama vector~\cite{kodama1980conserved}. 
Furthermore, the~Misner--Sharp mass in Equation~\eqref{general Misner--Sharp mass} leads to a generalized first law of black-hole thermodynamics that has its origins in~\cite{hayward1998unified,hayward2009local}
\begin{equation}
	\nabla_\mu M_\text{MS} = A \psi_\mu + w \nabla_\mu V~, 
	\label{first law}
\end{equation}
where $A$ and $V$ are the Euclidean area and volume of a 2-sphere, and~we assume that the relevant quantities can be expressed as functions of $(v,r)$. In~Equation~\eqref{first law}, the~energy-flux vector field ($\psi^\mu$) and work-density scalar ($w$) are given by
\begin{equation}
	\psi^\mu \equiv \frac{1}{8\pi}G^{\mu\gamma} \nabla_\gamma r + w \nabla^\mu r~, \hspace{1cm} w = - \frac{1}{16\pi} g_{ab} G^{ab}~, \label{gab}
\end{equation}
where $g_{ab}$ is the metric on the $(1+1)$-dimensional reduced spacetime orthogonal to the spherical orbits. In~the coordinates $(v,r)$,  
\begin{equation}
	\psi_\mu = \frac{1}{8\pi}G_\mu{}^r + w \delta_\mu{}^r~, \hspace{1cm} w = - \frac{G_v{}^v+G_r{}^r}{16\pi} ~,
\end{equation}
with the nonzero components of Equation~\eqref{first law}. Thus, 
\begin{equation}
	\partial_v M_\text{MS} = A \psi_v = \mathcal{A} ~, \hspace{1cm} \partial_r M_\text{MS} = 4\pi r^2\bigg(\frac{1}{8\pi}G_r{}^r + 2w\bigg) = \mathcal{B} ~. 
\end{equation}
This analysis shows that Equation~\eqref{general Misner--Sharp mass} is an integral form of Equation~\eqref{first law} for the Kodama vector $\text{K}^\mu = e^{-\lambda}(\partial_v)^\mu$.
Up to this point, there are no approximations~involved.

\section{Near-Horizon Approximation~Scheme}\label{sec:approx}

\subsection{Existence of Future Trapping~Horizons}
\label{sec:trap}

Following Hayward's approach, we define a black hole in terms of its trapping horizons. More precisely, we assume that the black-hole boundary is a future outer trapping horizon, a~concept that is properly reviewed in Appendix~\ref{app}.

Future outer trapping horizons and future inner trapping horizons  are characterized (respectively) by the conditions 
${\mathcal{B}(v, r_+(v)) < \nicefrac{1}{2} }$ and  ${ \mathcal{B}(v,r_-(v)) > \nicefrac{1}{2} }$ on the surface $r = 2M_\text{MS}$. 
The sign of $\mathcal{B}$, and~thus the type of horizon, depends on the component $T_v{}^v$ of the energy--momentum tensor and also on the sign and magnitude of the cosmological constant $\Lambda$. 
We initially consider the case of a vanishing $\Lambda$ and write $T_v{}^v = T^v{}_\mu \delta^\mu_v \,$.
We observe that $T_v{}^v < 0$  in regions where $\partial_v$ is future-directed, assuming the dominant energy condition holds. Hence $\mathcal{B} > 0$ within these assumptions,
with large (in module) $T_v{}^v$ contributing to the preferable existence of FITHs. Although~the positivity of $\mathcal{B}$ is assured by the dominant energy condition in this case, in~regions of $g_{vv} > 0$, the sign of $\mathcal{B}$ is not determined \textit{a priori}. 
If a cosmological constant is present, a~$\Lambda > 0$ also contributes to the formation of FITHs. But~that is not the case if $\Lambda < 0$.

Several physically significant solutions to the Einstein field equations exhibit specific energy conditions, $T_v{}^v$ negative and $\rho_\Lambda$ positive everywhere, which preferentially support the existence of FITHs~\cite{griffiths2009exact}. 
Prime examples are the Reissner--Nordstrom (low electric charge) and Schwarzschild-de Sitter black holes, including their dynamical generalizations~\cite{bonnor1970spherically,mallett1985radiating}.
In such spacetimes, the~standard configuration consists of concentric trapping horizons, with~both the FOTH and the FITH enclosing the central~singularity.

In an extremal case, FOTH and FITH merge onto a single trapping horizon. This condition is formally characterized by a vanishing geometric surface gravity, as~presented in Appendix~\ref{app}. 
A third, more radical, scenario arises when both horizons disappear completely, exposing a naked singularity. Such cases raise profound theoretical challenges,  as~they represent a violation of the cosmic censorship conjecture, which is beyond the scope of the present~work.

In summary, the~existence of black holes with multiple horizons requires specific conditions that ensure the simultaneous presence of both the outer and inner horizons. When the black hole is perturbed, the~positions of these horizons may shift, requiring a reexamination of the conditions that guarantee the existence of the black hole with a well-defined boundary. In~this work, we develop a systematic perturbative scheme for investigating accretion-induced shifts in the horizons of an accreting black~hole.

\subsection{General Model\label{sec:approx~VdS} }

Aiming at this dynamical analysis of the trapping horizons subjected to perturbations, an approximation scheme is developed in this section to simplify the metric near a surface $r = r_0$\,.
We have seen that, in~the appropriate coordinates of Equation~\eqref{vr vanish}, the~metric tends to a Vaidya form near $r_0$\,. 
Moreover, we will show that, up~to a first correction of the Misner--Sharp mass near 
$r_0$, the~metric locally acquires a form that can be related to the Vaidya--de Sitter spacetime.

For the subsequent analysis, consider the parameter $\epsilon$\,, defined as
\begin{equation}
	\epsilon \equiv r - r_0 ~ .
	\label{eps}
\end{equation}
We will approximate the Misner--Sharp mass around $r=r_0$\,, treating $\nicefrac{\epsilon}{r_0}$ as a small parameter.
Concretely, the~Vaidya--de Sitter line element,
\begin{equation}
	ds^2 = - \left\{ 1 - \frac{2}{r} \left[ m(v) + \frac{4 \pi r^3}{3} \rho_\Lambda \right] \right\} dv^2 + 2dvdr + r^2 d\Omega^2~,
\end{equation}
has its quasi-local mass approximated in the vicinity of $r_0$\,, $(r_0, r_0 + \epsilon)$\,, at~first order in $\nicefrac{\epsilon}{r_0} \,$, as
\begin{equation}
	M_\text{MS}(v,r) \approx M_\text{MS}(v,r_0) + 4\pi r_0^2 \epsilon \rho_\Lambda~,~~
	M_\text{MS}(v,r_0)=m(v)+ \frac{4\pi r_0^3}{3}\rho_\Lambda~.
	\label{eq:M-MS}
\end{equation}
The term $4\pi r_0^2 \epsilon$ corresponds to the (Euclidean) volume of a thin spherical shell of radius $r_{0}$ and thickness $\epsilon$. 
Thus, the~second term in the expansion of $M_\text{MS}(v,r)$ ($4\pi r_0^2 \epsilon \rho_\Lambda$) represents the energy contribution from the cosmological constant within this thin shell, providing a first-order correction to the Misner--Sharp mass.
Within this approximation, the~Vaidya--de Sitter geometry is characterized near $r_0$ by
\begin{equation}
	ds^2 = - \left\{ 1 - \frac{2}{r} \left[ m(v) + \frac{4 \pi r_0^3}{3} \rho_\Lambda + 4 \pi r_0^2 \epsilon \rho_\Lambda \right] \right\} dv^2 + 2dvdr + r^2 d\Omega^2~.
	\label{VdS around r0}
\end{equation}

More generally, using Equations~\eqref{general Misner--Sharp mass}--\eqref{MS mass (v,r)} and keeping only the leading-order contribution of the energy--momentum tensor in the region  $(r_0,r_0 + \epsilon)$\,, we obtain the following: 
\begin{equation}
	M_\text{MS}(v,r) \approx  M_\text{MS}(v,r_0) - 4\pi r_0^2 \epsilon [ T_v{}^v(v,r_0) -  \rho_\Lambda]~, 
	\label{VdS correction}
\end{equation}
with $M_\text{MS}(v,r_0)$ determined by initial conditions and the flow $\mathcal{A}$ across $r_0 \,$. 
This result implies that the metric can be approximated to a ``Vaidya-dark energy form'' around $r_0$\,, with~an effective dynamical cosmological constant $\tilde{\Lambda}$ given by
\begin{equation}
	\rho_{\tilde{\Lambda}}(v) \equiv -T_v{}^v(v,r_0) + \rho_\Lambda~.
	\label{effLambda}
\end{equation}
It follows that, given the condition
\begin{equation}
\epsilon \ll \left| \frac{\mathcal{B}(v,r_0)}{\mathcal{B}'(v,r_0)} \right|~,
\end{equation}
the Misner--Sharp mass in the vicinity of $r_0$ is approximated to the mass of a Vaidya geometry in a background dominated by a dynamical cosmological constant.

\subsection{Perfect Fluid~Model}\label{sec:approx perfect}

A particularly interesting case from our previous results concerns the accretion of perfect fluids. More specifically, let us consider an energy--momentum tensor
\begin{equation}
	T^{\mu\gamma} = (\rho + p)u^\mu u^\gamma + p g^{\mu\gamma}~,
	\label{eq:perfect-fluid}
\end{equation}
where the energy density $\rho$ and pressure $p$ are dynamical functions of both the time and radial coordinates, i.e.,~$\rho \equiv \rho(v,r)$ and $p \equiv p(v,r)$\,.

The 4-velocity $u^\mu$ of the fluid takes the form
\begin{equation}
	u^\mu = \big[F(v,r)\,, \ -G(v,r)\,,\ 0\,,\ 0 \big]~,
\end{equation}
with the radial velocity ($G$) required to be a positive function. 
The $v$-component ($F$) is determined by the normalization condition $u_\mu u^\mu = -1$\,,  yielding:
\begin{equation}
	u^{\mu} = \left( \frac{e^{-\lambda}}{G + \sqrt{G^2 + 1 - \frac{2M_\text{MS}}{r}}}\,, \ -G\,,\ 0\,,\ 0 \right)~.
	\label{u}
\end{equation}
This is the general form for the 4-velocity of a radially ingoing perfect fluid in a spherically symmetric spacetime using the coordinates of Equation~\eqref{general_metric}. Near~a fixed $r_0$, the~Misner--Sharp mass in Equation~\eqref{u} can be approximated by Equation~\eqref{VdS correction}.

Using the chart of Equation~\eqref{vr vanish}, the~relevant non-zero components of the energy--momentum tensor are written as
\begin{equation}
	T_v{}^{v} = \frac{-\rho \sqrt{1 - \frac{2M_\text{MS}}{r} + G^2} + p G}{\sqrt{1 - \frac{2M_\text{MS}}{r} + G^2} + G}~, 
	~~ 
	T_r{}^v= \frac{e^{-\lambda+\lambda_0}(\rho+p)}{\bigg[G + \sqrt{G^2 + 1 - \frac{2M_\text{MS}}{r}}\bigg]^2}~, 
	~~
	T_v{}^r = G (\rho + p) e^{\lambda-\lambda_0} \sqrt{1-\frac{2 M_\text{MS}}{r}+G^2}~,
	\label{Tmunu}
\end{equation}
with $\lambda_0 \equiv \lambda(v,r_0)$\,. 
From the previous development, 
the metric takes the form of a Vaidya-dark energy solution in the vicinity of the surface $r=r_{0}$\,, up~to first-order corrections of $\nicefrac{\epsilon}{r_0}$ on $M_\text{MS}$.
In the case of a perfect fluid, if~$r_0$ is the position of the trapping horizon at a particular $v$, then the effective vacuum energy $\rho_{\tilde{\Lambda}}$ in Equation~\eqref{effLambda} (associated to an effective dynamical cosmological constant $\tilde{\Lambda}$) acquires the form:
\begin{equation}
	\rho_{\tilde{\Lambda}}(v,r_0) = \frac{1}{2} [\rho(v,r_0) - p(v,r_0)] + \rho_\Lambda~. \label{rho perfect fluid}
\end{equation}

We emphasize that the near-horizon approximation,
\begin{equation}
	M_\text{MS}(v,r) \approx M_\text{MS}(v,r_0) + 2 \pi r_0^2 \epsilon \left[\rho(v,r_0) - p(v,r_0) + 2\rho_\Lambda \right]~,
	\label{VdS approx horizon}
\end{equation}
exhibits a leading-order correction determined solely by the physical scalars $\rho$, $p$ and $\rho_{\Lambda}$. That is, this correction term is independent of any explicit coupling between the fluid and the geometry, such as kinematic terms ($u^\mu$) or the metric itself ($g_{\mu\gamma}$).

\section{Perturbative Scheme for~Accretion}\label{sec:pert}  

\subsection{Perturbations near a Trapping~Horizon}\label{sec:pert near}

In the previous section, we developed a general approximation scheme for the Misner--Sharp mass, presented in Equation~\eqref{VdS correction}. This framework provides the specific form for a perfect fluid's energy--momentum tensor in Equation~\eqref{VdS approx horizon}, valid near a trapping horizon. Within~this scheme, the~surface $r = r_0$ represents the position of the horizon at a fixed value of the affine parameter, set as $v = 0$ for convenience.
To fully characterize the metric in this neighborhood of $r_0$\,, $M_\text{MS}(v,r_0)$ must be specified. 
However, this typically requires knowledge of the complete solution, which is generally not available.
In this context, a~perturbative approach proves particularly valuable, as~it allows us to consider fluid accretion on a fixed, predetermined black-hole~background.

The general framework of the perturbative scheme is given by a background metric $g_{\mu\gamma}^{(0)}$ that solves the Einstein Equation for a field $\phi^{(0)}$,
\begin{equation}
	G_{\mu\gamma}\left[g_{\mu\gamma}^{(0)} \right] = 8\pi T_{\mu\gamma}\left[g_{\mu\gamma}^{(0)} ~,\, \phi^{(0)} \right]~,
	\label{eq:equation-background}
\end{equation}
and the perturbation $g_{\mu\gamma}^{(1)}$ on the metric is produced by a field $\phi^{(1)}$ such that the Einstein Equation for the perturbed spacetime is given by
\begin{equation}
	G_{\mu\gamma}\left[g_{\mu\gamma}^{(0)} + g_{\mu\gamma}^{(1)}\right] = 8\pi T_{\mu\gamma}\left[g_{\mu\gamma}^{(0)} \, , \, \phi^{(0)} + \phi^{(1)}\right]~.
	\label{G pert}
\end{equation}
The energy--momentum tensor is determined by the test fluid approximation, requiring the matter fields on the right-hand side of Equation~\eqref{G pert} to satisfy their equations of motion on the background spacetime. Nonetheless, backreaction effects on the metric are considered, as~evidenced in the left-hand~side.

Our framework analyzes the backreaction on the background Misner--Sharp mass. While reference~\cite{babichev2012backreaction} established conditions for the validity of a similar perturbation scheme, we propose two significant refinements. First, the~fluid's dynamical degrees of freedom are preserved. Second, the~systematic near-horizon expansion in our method facilitates the computation of higher-order corrections. This approach provides a controlled approximation scheme for studying backreaction effects while maintaining consistency with the dynamical evolution of the~spacetime.

To analyze the behavior of the trapping horizon under a perturbation, we employ the Misner--Sharp mass decomposition within our perturbative framework,
\begin{equation}
M_\text{MS}(v,r) = M_\text{MS}^{(0)}(v,r) + M_\text{MS}^{(1)}(v,r)~, 
\label{M = M0 + M1}
\end{equation}
which requires
\begin{equation}
\frac{M_\text{MS}^{(1)}(v,r)}{M_\text{MS}^{(0)}(v,r)} \ll 1~.
\end{equation}
In Equation~\eqref{M = M0 + M1}, the~background and first-order contributions are given by the following:
\begin{align}
	&M_\text{MS}^{(0)}(v,r) = M_\text{MS}^{(0)}(0,r_0) + \int_0^v \mathcal{A}^{(0)} (v',r_0) dv' + \int_{r_0}^r \mathcal{B}^{(0)}(v,r') dr'~, \hspace{1cm} T_{\mu\gamma}^{(0)} = T_{\mu\gamma}\left[g_{\mu\gamma}^{(0)}, \phi^{(0)}\right]~, \label{pert1}
	\\
	&M_\text{MS}^{(1)}(v,r) = \int_0^v \mathcal{A}^{(1)} (v',r_0) dv' + \int_{r_0}^r \mathcal{B}^{(1)}(v,r') dr'~, \hspace{1cm} T_{\mu\gamma}^{(1)} = T_{\mu\gamma}\left[g_{\mu\gamma}^{(0)}, \phi^{(1)}\right]~. \label{pert2}
\end{align}
We have assumed that the accretion process begins strictly for $v > 0$, setting
\begin{equation}
	M_\text{MS}^{(1)}(0,r_0) = 0~. \label{M1(0,0)}
\end{equation}
That is, at~$v=0$\,, the~Misner--Sharp mass evaluated at the initial position of the trapping horizon $(r_0)$ is determined by the background alone. 

A natural initial application for the perturbative scheme involves a Schwarzschild background~\cite{griffiths2009exact}, with~mass parameter $m$ and $\phi^{(0)} = 0$. 
In this case, the~Misner--Sharp mass decomposes as
\begin{equation}
	M_\text{MS}^{(0)} (v,r) = M_\text{MS}^{(0)} = m~, ~~ 
	M_\text{MS}(v,r) = m +  \int_0^v \mathcal{A}^{(1)} (v',r_0) dv' + \int_{r_0}^r \mathcal{B}^{(1)} (v,r')dr'~.
\end{equation}

As a more complex scenario, we consider a Reissner--Nordstr\"om background~\cite{griffiths2009exact}, with~electric charge $Q$ and $\phi^{(0)}$ representing the electric potential.
In this case, the~Misner--Sharp mass can be written as follows:
\begin{align}
	M_\text{MS}^{(0)} (v,r) = M_\text{MS}^{(0)} (r)  = m - \frac{Q^2}{2r}~, ~~
	M_\text{MS} (v,r) = m - \frac{Q^2}{2r} +\int_0^v \mathcal{A}^{(1)} (v',r_0) dv' + \int_{r_0}^r \mathcal{B}^{(1)} (v,r')dr'~. 
	\label{RN pert}
\end{align}
In the following section, we examine how accretion affects the trapping horizons of a Reissner--Nordstr\"{o}m black~hole.

\subsection{Accretion-Induced Shifts in Trapping~Horizons} 
\label{sec:pert shift}

To incorporate the approximation scheme of Section~\ref{sec:approx} into the perturbative framework, we perform a systematic near-horizon expansion in powers of $\epsilon$, defined in Equation~\eqref{eps}, analyzing three successive orders:
\\
\textit{(Zeroth-order accreting solution)} The exact solution at $\epsilon = 0$ from Equation~\eqref{pert1}, including metric perturbations at $\epsilon = 0$ from Equation~\eqref{pert2}:
\begin{equation}
	M_{\text{MS}}^{[0]}(v,r) \equiv M_{\text{MS}}^{(0)}(v,r_0) + \bar{\mathcal{A}}^{(1)}(v,r_0)~. \label{ap1}
\end{equation}
\\
\textit{(First-order accreting solution)} The full exact solution from Equation~\eqref{pert1}, including metric perturbations at $\epsilon = 0$ from Equation~\eqref{pert2}:
\begin{equation}
	M_{\text{MS}}^{[1]}(v,r) \equiv M_{\text{MS}}^{(0)}(v,r) + \bar{\mathcal{A}}^{(1)}(v,r_0)~. \label{ap2}
\end{equation}
\\
\textit{(Full perturbation)} Incorporating the contribution $\mathcal{B}^{(1)}$ evaluated at $r_0$:
\begin{equation}
	M_{\text{MS}}^{[2]}(v,r) \equiv M_{\text{MS}}^{(0)}(v,r) + \bar{\mathcal{A}}^{(1)}(v,r_0) + \epsilon \mathcal{B}^{(1)}(v,r_0)~. \label{ap3}
\end{equation}
In Equations~\eqref{ap1}--\eqref{ap3}, we have defined
\begin{equation}
	\bar{\mathcal{A}}^{(1)}(v,r_0) \equiv \int_0^v \mathcal{A}^{(1)}(v',r_0)dv'~.
\end{equation}

Let $r_0$ denote the initial position of a trapping horizon. In~the unperturbed Reissner--Nordstr\"{o}m spacetime, its position is determined by
\begin{equation}
	r_0 = m \pm \sqrt{m^2 - Q^2}~, \label{RN r0}
\end{equation}
with the usual extremal condition:
\begin{equation}
	m \overset{\text{ext}}{=} |Q|~, \hspace{1cm} r_0 \overset{\text{ext}}{=}|Q|~. \label{RN ext}
\end{equation}
In what follows, we investigate small horizon shifts $\epsilon(v) \equiv r_H(v) - r_0$ induced by matter~accretion. 

Using the zeroth-order accreting solution of Equation~\eqref{ap1}, the~perturbed Misner--Sharp mass takes the form:
\begin{equation}
	M_\text{MS}^{[0]}(v, r) = m + \bar{\mathcal{A}}^{(1)}(v,r_0) - \frac{Q^2}{2r_0}~.
\end{equation}
This leads to a time-dependent horizon shift:
\begin{equation}
	\frac{r_H^{[0]}(v)}{2} = m - \frac{Q^2}{2r_0} + \bar{\mathcal{A}}^{(1)}(v,r_0)~.
	\label{r_H charged (0)}
\end{equation}

Extending to our first-order accreting solution of Equation~\eqref{ap2}, the~Misner--Sharp mass becomes
\begin{equation}
	M_\text{MS}^{[1]}(v, r) = m + \bar{\mathcal{A}}^{(1)}(v,r_0) - \frac{Q^2}{2r}~,
\end{equation} 
which generates shifts in the horizon as
\begin{equation}
	\frac{r_H^{[1]}(v)}{2} = m - \frac{Q^2}{2 r_H^{[1]}(v)} + \bar{\mathcal{A}}^{(1)} (v,r_0) ~.
	\label{eq:rational_rH}
\end{equation}
This rational equation~\eqref{eq:rational_rH} in $r_H^{[1]}(v)$ yields the exact solution:
\begin{equation}
	r_H^{[1]}(v) = m + \bar{\mathcal{A}}^{(1)} (v,r_0) \pm \sqrt{ \left[\bar{\mathcal{A}}^{(1)} (v,r_0)\right]^2 + m^2 - Q^2 + 2 m \bar{\mathcal{A}}^{(1)} (v,r_0)}~.
	\label{r_H charged (1)}
\end{equation}
Note that, in~this first-order approximation, the~horizon position remains fixed for $v\leq 0$ as a consequence of assumption~\eqref{M1(0,0)}, reducing to Equation~\eqref{RN r0}.

At this order of approximation, the~extremal condition is given by 
\begin{equation}
	m \overset{\text{ext}}{=} |Q| - \bar{\mathcal{A}}^{(1)} (v,r_0)~, \label{first ext cond}
\end{equation}
with the corresponding horizon position obtained from Equation~\eqref{r_H charged (1)},
\begin{equation}
	r_H^{[1]} (v) \overset{\text{ext}}{=} |Q|~.
\end{equation}
Thus, it is found that the corrected condition for the existence of the trapping horizons is
\begin{equation}
	\bar{\mathcal{A}}^{(1)} (v,r_0) \geq -m + |Q| ~,
	\label{cond A}
\end{equation}
for all values of $v$. 

Since the initial condition $M_{\text{MS}}(0,r_0) = M_{\text{MS}}^{(0)}(0,r_0)$ implies $m \geq |Q|$, the~inequality \eqref{cond A} is automatically satisfied for physical fluids obeying the traditional energy conditions. This suggests that matter influx alone can only drive an extremal horizon to split into distinct inner and outer horizons. 
That is, the~influx of electrically neutral perfect fluid directly leads to the breakdown of the extremality condition~\eqref{first ext cond}.
Furthermore, it also indicates that phantom-energy accretion (with $\bar{\mathcal{A}}^{(1)} < 0$) could violate this condition, leading to horizon~disappearance.

So far, our analysis has not considered the contribution of the energy density. To~include it, we use the full perturbation of Equation~\eqref{ap3}:
\begin{equation}
	M_\text{MS}^{[2]}(v,r) = m + \bar{\mathcal{A}}^{(1)}(v,r_0) - \frac{Q^2}{2r} + \epsilon \mathcal{B}^{(1)}(v,r_0)~. 
\end{equation}
For a dust fluid, generalization discussed previously as $\rho_{\tilde{\Lambda}}$, the~horizon position becomes the following:
\begin{equation}
	\frac{r_H^{[2]}(v)}{2} =  m - \frac{Q^2}{2r_H^{[2]}(v)} + \bar{\mathcal{A}}^{(1)}(v,r_0) + 2 \pi r_0^2 \left[r_H^{[2]}(v)-r_0\right] \rho(v,r_0)~.
	\label{eq:rH2}
\end{equation}
The exact solution to Equation~\eqref{eq:rH2} is
\begin{equation}
	r_H^{[2]}(v) = \frac{2 \pi r_0^3 \rho(v,r_0) -\bar{\mathcal{A}}^{(1)}(v,r_0) - m \pm \sqrt{Q^2 \left[4 \pi r_0^2 \rho(v,r_0) -1\right]+\left[ \bar{\mathcal{A}}^{(1)}(v,r_0) + m - 2 \pi r_0^3 \rho(v,r_0) \right]^2} }{4\pi r_0^2 \rho(v,r_0) - 1}~.
	\label{r_H charged (2)}
\end{equation}

The corrected extremal condition now incorporates energy density effects:
\begin{equation}
	m \overset{\text{ext}}{=} - \bar{\mathcal{A}}^{(1)}(v,r_0) + 2 \pi r_0^3 \rho(v,r_0)+ \sqrt{Q^2\left[1- 4 \pi r_0^2 \rho(v,r_0)\right]}~,
\end{equation}
with the position of the extremal horizon given by
\begin{equation}
	r_H^{[2]} (v) \overset{\text{ext}}{=} \frac{|Q|}{\sqrt{1 - A_0 \rho(v,r_0) }}~, \hspace{1cm} A_0 = 4\pi r_0^2~.
	\label{r_H ext first order}
\end{equation}
The framework is reliable for small perturbations, $\rho(v,r) \sim 0$. Hence, up~to a first correction, the~position of the trapping horizon of a perturbed extremal Reissner--Nordstr\"om black hole is given by
\begin{equation}
	r_H^{[2]} (v) \overset{\text{ext}}{=} |Q|\left[ 1 + \frac{A_0 \rho(v,r_0)}{2} \right]~. 
\end{equation}

Some key advances of our perturbative approach are discussed. The~result~\eqref{r_H charged (1)} shows that retaining second-order influx terms in the square root induces first-order corrections to the extremal condition, leading to the necessity of a revised analysis of the existence criteria for trapping horizons. Also, result~\eqref{r_H charged (2)} incorporates energy density effects, generating a corrected extremal condition with additional terms, as well as a shifted horizon position for extremal Reissner--Nordstr\"om black holes.
Finally, our perturbative approach produced higher order corrections than those of~\cite{babichev2012backreaction}. 

This section presents the main results of our work. 
However, we further examine the properties of future inner trapping horizons, given their profound theoretical~implications.

\section{On the Presence of Future Inner Trapping~Horizons} \label{sec:FITH}

As established in Section~\ref{sec:trap}, future inner trapping horizons are found in many solutions. However, their presence implies that the standard black-hole boundary definition via a future outer trapping horizon must now be complemented with additional existence conditions. This  requires a particularly careful treatment when studying perturbations on extremal and near-extremal configurations, as was carried out in the previous~section.

The existence of inner horizons also leads to a variety of potential theoretical challenges: as possible Cauchy horizons, raising questions regarding the predictability of general relativity~\cite{penrose1968structure}; and as unstable loci~\cite{matzner1979instability}, most notably resulting in the mass inflation phenomenon~\cite{poisson1990internal,ori1991inner,di2022inner}. Recent work has reinterpreted these phenomena for FITHs in dynamical spacetimes~\cite{carballo2024mass}, moving beyond the traditional analysis based on stationary black~holes.

It is thus remarkable that we can identify small metric corrections that incorporate repulsive effects, which can eliminate the inner horizon in some scenarios. This mechanism generates a secondary FOTH within the FITH. We have observed that the smaller the FITH is, the~easier it is to destroy it. For~sufficiently strong corrections, this inner FOTH can completely cancel out the FITH, leaving only the outer horizon as the boundary of the black hole.
More precisely, we consider specific deviations of the form
\begin{equation}
	M_\text{MS} \longrightarrow M_\text{MS} + \alpha r^{-\beta}~,
\end{equation}
where $\alpha$ is positive and $\beta$ is a non-negative integer.
These corrections are well motivated by several works in the literature, particularly spacetimes incorporating quantum gravity effects.
For instance, similar terms emerge as one-loop corrections in gravitational effective field theories (see~\cite{bargueno2016quantum} and references therein).
As a concrete scenario, the~FOTH of a Schwarzschild can be shifted (but never destroyed) for any pair of $(\alpha,\beta)$. For~a Reissner--Nordstrom black hole,
\begin{equation}
	M_\text{MS} = m - \frac{Q^2}{2r} \longrightarrow m - \frac{Q^2}{2r} + \alpha r^{-\beta}~.
\end{equation}

\begin{figure}[h]
	\includegraphics[width=0.5\textwidth]{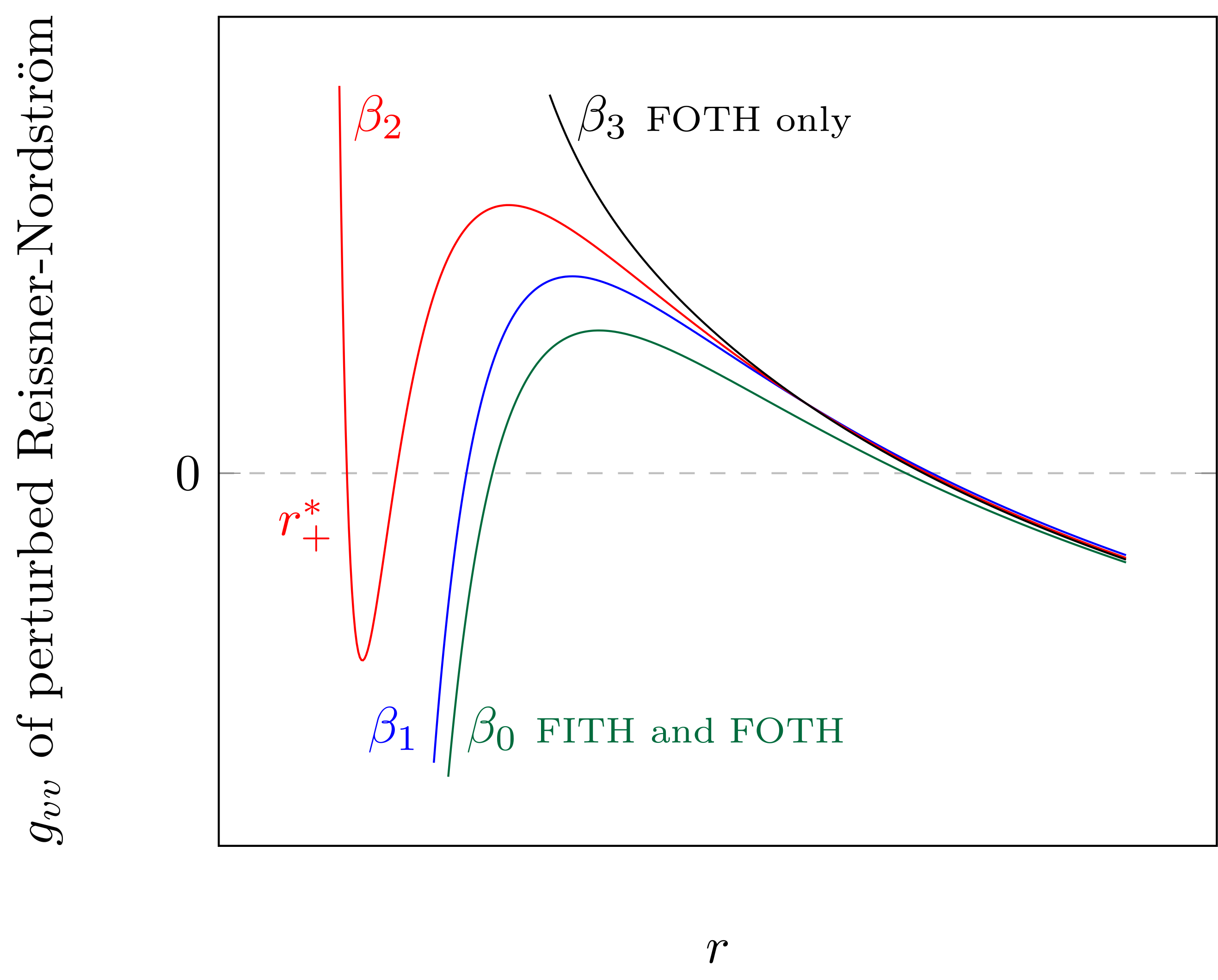}
	\caption{General representation of the metric function $g_{vv}$ of a pure Reissner--Nordstr\"om black hole $(\beta_0)$ with (FITH) $r_- \sim 0$ corrected by the same $\alpha$ and increasing values of $\beta$: from $\beta_1$ to $\beta_3$. The~quantity $r_+^*$ denotes a secondary FOTH that is formed between the singularity and the FITH by the correction $\beta_2$.}
	\label{fig1}
\end{figure}

As shown in Figure~\ref{fig1}, there are
qualitatively different repulsive corrections on the metric of a Reissner--Nordstr\"om black hole, which favors the vanishing of the FITH for increasing values of $\beta$. In~fact, virtually undetectable corrections to the metric by an exterior observer can be sufficient to make the inner horizon~vanish.

The perturbative analysis developed in the previous section can be straightforwardly extended to Reissner--Nordstr\"om with this correction. For~several values of the parameters $\alpha$ and $\beta$, the~solution admits a single (outer) trapping horizon, defining its boundary. This simplifies the perturbative treatment and avoids potential instabilities arising from the presence of the inner~horizon. 

\section{Final~Remarks}\label{sec:final}

In this work, we have investigated the dynamics of future trapping horizons of spherically symmetric spacetimes using ingoing Eddington--Finkelstein coordinates. We have shown that their exact evolution is described by an integral form of Hayward's first law.
Our results demonstrate the important role of quasi-local quantities, in~particular trapping horizons and the Misner--Sharp mass, in~characterizing the accretion-induced evolution of black holes. These tools are particularly valuable when exact solutions incorporating backreaction effects are not available and~when going beyond standard test-fluid approximations is~desirable.

To consider backreaction in our analysis, we have implemented a near-horizon approximation scheme that systematically captures first-order corrections of the Vaidya-dark energy type. This framework is particularly useful for simplifying the analysis of perfect fluids near trapping horizons since the leading-order corrections depend only on the fluid's energy density $\rho$ and pressure $p$, with~no explicit dependence on either the fluid's velocity field or the background metric components. This simplification arises because, in~the first correction, the~interaction of the fluid with the geometry is fully encoded in the stress-energy tensor via $\rho$ and $p$.

Building on this foundation, our perturbative framework extends previous results by systematically incorporating higher-order corrections. The~formalism is constructed in a background-independent manner and subsequently applied to study accretion onto a Reissner--Nordstr\"{o}m black hole.
Our analysis reveals a hierarchy of horizon displacements, with~distinct perturbative orders producing characteristically different modifications to both the horizon location and the extremality conditions. Significantly, we find that momentum influx (from non-exotic matter) preserves the coordinate location of an extremal horizon, serving only to bifurcate it into separate inner and outer components. Energy density contributions, by~contrast, produce measurable shifts of the extremal horizon~position.

The generality of our framework, combined with its systematic perturbative construction, opens up several promising directions for future research. We have observed that certain repulsive corrections in the metric are able to eliminate an inner horizon, thus avoiding the additional theoretical challenges they pose. It would be of significant interest to investigate this effect within a perturbative framework. However, such an analysis would likely require going beyond the localized perturbations used in accretion models.
Furthermore, since these corrections directly impact the Misner--Sharp mass, they yield new results for quasi-local versions of black-hole thermodynamics. This suggests the need for further investigation into the effects of metric corrections on horizon~structure.

Moreover, it is possible to implement our methodology to outgoing Eddington--Finkelstein coordinates, which can be used to model Hawking evaporation~\cite{hiscock1981modelsI,hiscock1981modelsII,campos2024}. Also, in~this case, the~formalism could be applied to the study of the dynamics of past cosmological horizons, which are those of fast-expanding cosmologies. 
Since the perturbative approach developed here is valid close to a trapping horizon, it naturally enables the investigation of dynamical configurations in near-extremal regimes.  This is particularly true for spacetimes like Vaidya--de Sitter and its generalizations.
Another extension of this work involves incorporating higher-order perturbations, for~which a better formalization of our theoretical framework is required.
Research is being conducted along these~lines.

\appendix
\section{Future Trapping~Horizons}\label{app}

The boundary of a black hole is considered in this work to be a future outer trapping horizon (FOTH), which is characterized by the conditions that are made explicit in this appendix. The~brief presentation starts with the expansion scalar, which measures the rate of change of the transverse metric $h_{\mu\gamma}$ along the generators $k$ of a null%
\footnote{The investigation of null congruences is useful for the study of trapping horizons, but~the expansion scalar can also be defined for timelike congruences.}
congruence,
\begin{equation}
	\theta_k = \frac{1}{\sqrt{h}} \partial_k \sqrt{h}~, \label{definition of expansion scalar}
\end{equation}
where $h$ denotes the determinant of the transverse metric $h_{\mu\gamma}$:
\begin{equation}
	h \equiv \det \left( h_{\mu\gamma} \right) ~, ~~
	h_{\mu\gamma} \equiv g_{\mu\gamma} + l_\mu n_\gamma + n_\mu l_\gamma~. 
	\label{definition h}
\end{equation}
The tensor $h_{\mu\gamma}$ projects vectors on the cross-sectional surface orthogonal to the null vector fields $l^\mu$ and $n^\mu$. 

In this work, we consider the metric of a spherically symmetric geometry in the general form of Equation~\eqref{general_metric}. For~the null fields, we take the ingoing and outgoing vectors:
\begin{align}
	&n = -\partial_r~,
	\\
	&l = e^{-\lambda}\partial_v + \frac{1}{2} \left( 1 - \frac{2 M_\text{MS}}{r} \right) \partial_r~. 
\end{align}
It can be checked from Equation~\eqref{definition h} that the area element of the cross-sectional surface orthogonal to both $l^\mu$ and $n^\mu$ is
\begin{equation}
	\sqrt{h} = r^2 \sin\theta~,
\end{equation}
which characterizes a 2-sphere. And, for~the expansion scalar~\eqref{definition of expansion scalar},
\begin{align}
	&\theta_n = -2r\sin \theta~,
	\\
	&\theta_l = \frac{1}{r} \left( 1 - \frac{2M_\text{MS}}{r} \right) ~.
\end{align}

By definition~\eqref{def th}, there is a future trapping horizon at $r_H(v)$ if
\begin{equation}
	r_H(v) = 2M_\text{MS}~.
\end{equation} 
This structure is characterized by the sign of $\mathcal{L}_n \theta_l$ at $r_H$, where $\mathcal{L}$ denotes the Lie derivative.
For a FOTH we have, for~all $v$,
\begin{equation}
	\mathcal{L}_n \theta_l \big|_{r_H(v)} < 0 
	~~ 
	\iff 
	~~ 
	\mathcal{B} \left( v, r_H(v) \right) \equiv \frac{\partial M_\text{MS}}{\partial r} \bigg|_{r_H(v)} < \frac{1}{2}~.
\end{equation}
And for it to be a future inner trapping horizon (FITH), which is a typical boundary of a contracting spacetime,
\begin{equation}
	\mathcal{L}_n \theta_l \big|_{r_H(v)} > 0 
	~~ 
	\iff 
	~~ 
	\mathcal{B} \left( v,r_H(v) \right) \equiv \frac{\partial M_\text{MS}}{\partial r} \bigg|_{r_H(v)} > \frac{1}{2}~.
\end{equation}

When $\mathcal{B} = \nicefrac{1}{2}$, which suggests FOTH and FITH merging, we have an extremal trapping horizon. This is equivalently characterized by the vanishing of the geometric surface gravity:
\begin{equation}
	\kappa_G \equiv \frac{1}{2\sqrt{\text{det}|g_{ab}|}} \partial_c \left(\sqrt{\text{det}|g_{ab}|} \, g^{cd}\partial_d r\right) = - \partial_r \left(\frac{M_\text{MS}}{r} \right)~,
\end{equation}
with $g_{ab}$ defined after Equation~\eqref{gab}, and~the second equality given by evaluation in the coordinates of Equation~\eqref{general_metric}, where Latin indexes stand for $v$ and $r$.

There are two other types of trapping horizons: the~past (outer and inner) ones. However, these are covered by outgoing Eddington--Finkelstein coordinates, which are not as useful for investigating the dynamics of black-hole~accretion.

\section*{Funding}

T.~L.~C. acknowledges the support of Coordena\c{c}\~ao de Aperfei\c{c}oamento de Pessoal de N\'{\i}vel Superior (CAPES) -- Brazil, Finance Code 001.
C.~M. is supported by Grant No.~2022/07534-0, S\~ao Paulo Research Foundation (FAPESP), Brazil.

\begin{acknowledgments}
	
M. C. B. thanks Professor Orfeu Bertolami at University of Porto for hosting part of this study.
	
\end{acknowledgments}

\end{document}